\documentclass[letterpaper, 10pt, conference]{ieeeconf}  
\IEEEoverridecommandlockouts
\usepackage{cite}
\usepackage{graphicx}
\usepackage{array}
\usepackage{hyperref}
\usepackage{arydshln}
\usepackage{float}
\usepackage{scrextend}
\usepackage{blindtext}
\usepackage{hyphenat}
\usepackage{xspace}

\let\labelindent\relax

\makeatletter
\g@addto@macro\normalsize{%
  \setlength\abovedisplayskip{3.2pt plus 1.3pt minus .8pt}
  \setlength\belowdisplayskip{\abovedisplayskip}
  \setlength\abovedisplayshortskip{.5pt plus 1pt}
  \setlength\belowdisplayshortskip{\abovedisplayshortskip}
}
\let\OLDthebibliography\thebibliography
\renewcommand\thebibliography[1]{
  \OLDthebibliography{#1}\vspace{-1.5mm}
  \setlength{\parskip}{0pt}
  \setlength{\itemsep}{0pt plus 0.1ex}
}

\newcommand{\shortsubsection}[1]{\vspace{-1mm}\subsection{#1}\vspace{-1.0mm}}
\newcommand{\shortsection}[1]{\vspace{-.5mm}\section{#1}\vspace{-1.5mm}}

\usepackage{tikz}
\usepackage{xspace}
\usepackage{xcolor}
\usepackage{amsthm}
\usepackage{amsmath,amssymb,amsfonts}
\usepackage{mathrsfs,mathtools}
\usepackage{mathalfa}
\usepackage{euscript}
\usepackage{csquotes}
\usepackage{enumitem}
\usepackage[normalem]{ulem}

\definecolor{mblue}{rgb}{0,0.4470,0.7410}
\definecolor{morange}{rgb}{0.8500,0.3250,0.0980}
\definecolor{myellow}{rgb}{0.9290,0.6940,0.1250}
\definecolor{mpurple}{rgb}{0.4940,0.1840,0.5560}
\definecolor{mgreen}{rgb}{0.4660,0.6740,0.1880}
\definecolor{mcyan}{rgb}{0.3010,0.7450,0.9330}
\definecolor{mred}{rgb}{0.6350,0.0780,0.1840}
\definecolor{mgreenblue}{rgb}{0.0,1.0,0.5}
\definecolor{parulablue}{rgb}{0.2431,0.1490,0.6588}
\definecolor{parulalblue}{RGB}{39,151,235}
\definecolor{parulagreen}{RGB}{129,204,89}
\definecolor{parulayellow}{RGB}{249,251,21}

\definecolor{cblue}{rgb}{0,0.9,1}
\definecolor{corange}{rgb}{1,0.7,0}

\theoremstyle{definition}
\newtheorem{defn}{Definition}
\newtheorem{exmp}{Example}
\theoremstyle{plain}

\theoremstyle{remark}

\newcommand{\lpvcore}{\textsc{LPVcore}\xspace}
\newcommand{\matlab}{\textsc{Matlab}\xspace}
\newcommand{\comment}[1]{}

\newcounter{ass}

\newcommand{\mc}[1]{\mathcal{#1}}

\newcommand{\mr}[1]{\mathrm{#1}}
\newcommand{\mb}[1]{\mathbb{#1}}

\DeclareFontFamily{U}{txcal}{\skewchar \font =45}
\DeclareFontShape{U}{txcal}{m}{n}{<-> txr-cal}{}
\DeclareMathAlphabet{\mathcalpxtx}{U}{txcal}{m}{n}

\newcommand{\chris}[1]{#1}

\newcommand{\dny}{n_\mr{y}}

\newcommand{\dnp}{n_\mr{p}}

\newcommand{\dna}{n_\mr{a}}
\newcommand{\dnb}{n_\mr{b}}

\def\methodname{LPV-SUBNET\xspace}

\title{Deep-Learning-Based Identification of LPV Models for Nonlinear Systems}

\author{Chris Verhoek, Gerben I. Beintema, Sofie Haesaert, Maarten Schoukens, and Roland T\'oth
\thanks{\chris{This work was supported by the European Space Agency in the scope of the `AI4GNC' project with SENER Aeroespacial S.A. (contract nr. 4000133595/20/NL/CRS) and was also supported by the European Union within the framework of the National Laboratory for Autonomous Systems (RRF-2.3.1-21-2022-00002).}}%
\thanks{The authors
are with the Control Systems Group, Eindhoven University of Technology, The Netherlands. R. T\'oth is also with the Institute for Computer Science and Control, Budapest, Hungary. 
Corresponding author: C. Verhoek (\texttt{c.verhoek@tue.nl}).}
}

\begin{document}
\maketitle
\thispagestyle{empty}
\pagestyle{empty}

\begin{abstract}
%
%
The Linear Parameter-Varying (LPV) framework provides a modeling and control design toolchain to address nonlinear (NL) system behavior via linear surrogate models. Despite major research effort on LPV data-driven modeling, a key shortcoming of the current identification theory is that often the scheduling variable is assumed to be a given measured signal in the data set. In case of identifying an LPV model of a NL system, the selection of the scheduling map, which describes the relation to the measurable scheduling signal, is put on the users' shoulder, with only limited supporting tools available. This choice however greatly affects the usability and complexity of the resulting LPV model. This paper presents a deep-learning-based approach to provide \emph{joint} estimation of a scheduling map and an LPV state-space model of a NL system from input-output data, and has consistency guarantees under general innovation-type noise conditions. Its efficiency is demonstrated on a realistic identification problem.
\end{abstract}
\shortsection{Introduction}\label{sec:introduction}
\noindent With the continuous push for increasing performance and energy efficiency of systems in engineering, like high-tech mechatronic devices, air and spacecrafts, power systems, etc., \emph{nonlinear} (NL) and \emph{time-varying} (TV) effects have become dominant in the dynamic behavior 
of these systems. To cope with these effects in engineering, an efficient modeling and control design toolchain is required that is capable of building on the already existing, vast experience with  \emph{Linear Time-Invariant} (LTI) design tools.  The \emph{Linear Parameter-Varying} (LPV) framework
has been established for this purpose, providing accurate, but low-complexity surrogate representations of NL and TV systems \cite{Shamma92}. In LPV systems, the signal relations are considered to be linear, just as in the LTI  case, but
the parameters defining these relations are assumed to be functions of a measurable,
time-varying signal, the so-called \emph{scheduling variable}
$p:\mathbb{Z}\rightarrow\mathbb{P}\subseteq
\mathbb{R}^{n_{\mathrm{p}}}$, which captures the original NL/TV effects in the system \cite{ref:mybook}. The LPV system class has shown to be capable of representing a wide variety of physical processes, but its major advantage is its well-worked-out and industrially-reputed
controller design framework, which allows to guarantee closed-loop 
stability and performance of the original physical system.

In order to support the use of LPV control design methods, a serious research effort has been spent on deriving a wide range of NL model conversion approaches 
to transform existing first-principle models of the system to low-complexity LPV models, see e.g. \cite{Hoffmann2015b,Toth20bIET,ref:mybook,Casella2009}. To formulate the \chris{LPV} signal relations, these methods should provide (i) construction of a \emph{scheduling map} $\phi: \mathbb{R}^n \rightarrow \mathbb{P}$ that describes how external or internal signals of the system are mapped to a scheduling variable $p$, and (ii) derive $p$-dependent \emph{coefficient functions}, like mappings $A,B,C,D$ in case of an LPV \emph{State-Space} (SS) model \cite{ref:mybook}. In general, low complexity of the LPV signal relations is preferred to simplify the follow-up utilization of the resulting model. Low complexity is often interpreted in terms of, e.g., affine scheduling dependence of the coefficient functions and a minimal scheduling dimension $\dnp$. Furthermore, deploying LPV controllers that are designed based on such models require real-time computation of $p$. Hence, scheduling variables that can be directly computed from measurable signals coming from the system, without the need of NL observers, are highly preferable. 
In overall, it has been realized that selection of $\phi$ has paramount importance in the conversion process, as it governs the expected performance and utilization potential of the overall LPV toolchain \cite{Hoffmann2015b}. 

However, \chris{(LPV)} models of the considered engineering systems that have the desired level of accuracy are often not available in practice. Moreover, it is usually difficult to decide which specific aspects of first-principles models are essential for representing the dynamic behavior of the system. {Hence, a wide range of data-driven modeling methods have been introduced that aim to estimate LPV models directly from measured \emph{input-output} (IO) data of the system.} Methods ranging from \emph{continuous} to \emph{discrete time}, using various model structures, such as, SS, IO, \emph{series expansion} and \emph{linear fractional} representations, under a wide range of noise scenarios, and using only \emph{local} operating-point-based measurements, \emph{global} measurements or both, have been developed and successfully applied in practice, see the overviews in \cite{ref:mybook,LPVbook2,Toth14JPC}. Especially \emph{prediction error minimization} (PEM) methods \cite{toth2012prediction, zhao2012prediction} and \emph{sub-space} techniques \cite{Cox2021,Wingerden09} have proven to be quite successful, while \emph{sparsity}-based model structure selection tools \cite{Toth12cCDC} and learning methods, such as, \emph{support vector machines} 
\cite{Toth18bAUT,Toth19bAUT,DOSSANTOS20197}, \emph{Gaussian processes} \cite{Toth18aAUT} and \chris{\emph{Artificial Neural Network} (ANN)} methods \cite{Lachhab2008,BAO20205286,ref5717855,ref6967628}, have also been introduced to provide flexible estimators of the required coefficient functions dependency. However, these methods almost exclusively rely on a \emph{predetermined} choice of $p$, assuming that its variation is part of the measured data set.  This puts the selection of the scheduling map $\phi$ on the users' shoulder, although it is such an important part of the overall modeling process. So far, a data-based choice for the scheduling map is only developed for a limited set of LPV-IO PEM methods by using sparse regression-based methods. These, however, still rely on a predefined set of signals and relations on which this choice is based. 
This difficulty comes from the overall complexity of the joint estimation problem of $\phi$ together with the $p$-dependent coefficient functions, constituting the LPV model. This joint estimation problem is inherently a \emph{NL identification} problem, requiring flexible function estimators, without restricting \mbox{a priori} choices and constraints. 

To overcome this challenging problem, we present a PEM-based deep-learning approach to provide joint estimation of a low-complexity scheduling map \emph{and} an affinely dependent LPV-SS model, \chris{to which we refer to as the \emph{joint LPV identification problem}. Our approach} allows to accurately identify a NL system directly from IO data, without any structural knowledge or a priori choices {required from} the user. 
The presented method builds on recent results in NL system identification in terms of the \emph{Sub-Space Encoder Network}  (SUBNET) method \chris{\cite{beintema2021nonlinear, beintema2021nonlinearvideo, Toth22a}}. SUBNET uses a novel state-encoder and a batch-wise formulation of the prediction cost, which results in an efficient NL identification approach. Our contribution in this paper is the modification of SUBNET to the joint LPV identification problem, under innovation-type noise structures, resulting in an efficient estimator for NL systems. The provided method is able to estimate LPV models for which the internal state of the model itself is also used to determine the scheduling, i.e., the model is capable of \emph{self-scheduling}. Furthermore, we introduce an alternative formulation to estimate LPV models for which the scheduling is \emph{directly} computed from past IO data. This alternative formulation yields a practically implementable scheduling map for controllers that are designed for the resulting model.  The approach has consistency guarantees under general innovation-type noise conditions and its efficiency is demonstrated on the identification problem of a control moment gyroscope. Note that, compared to other neural network based LPV identification methods (e.g., \cite{Lachhab2008,BAO20205286,ref5717855,ref6967628}), this is by our knowledge the first approach where (i) $p$ is \emph{not} part of the data set, (ii) the corresponding scheduling map $\phi$ is directly learned from past IO data, and (iii) a noise model is estimated together with the process model. 


\chris{Section \ref{sec:problemstatement} introduces the joint LPV identification problem for NL systems, while the proposed LPV sub-space encoder approach is discussed in Section \ref{sec:preliminaries}. We demonstrate the strength of the approach on a simulation study in Section \ref{sec:examples} and the conclusions are given in Section \ref{sec:conclusion}.}

\shortsection{Problem statement}\label{sec:problemstatement}
\noindent Consider the \emph{discrete-time} (DT) data-generating system given by the NL SS representation
\begin{equation}\label{eq:datagen_sys}
\Sigma : \left\{\begin{aligned} \quad x_{k+1} & = f(x_k,u_k,w_k), \\ \quad y_k & = h(x_k,u_k)+w_k, \end{aligned} \right.
\end{equation}
where $x_k\in\mb{X}\subseteq \mb{R}^{n_\mr{x}}$ is the state, $u_k\in\mb{U} \subseteq \mb{R}^{n_\mr{u}}$ is the input and $y_k\in\mb{Y} \subseteq  \mb{R}^{n_\mr{y}}$ is the observed output at time moment $k\in\mathbb{Z}$, while $w_k\in\mb{W}\subseteq\mb{R}^{n_\mr{y}}$ is the realization of an i.i.d.~white noise process with finite variance $\Gamma_\mathrm{w} \in \mb{R}^{n_\mr{y}\times n_\mr{y}}$. Moreover, $\mb{X}$, $\mb{U}$, $\mb{Y}$ and $\mb{W}$ are considered to be open sets containing the origin, and $f: \mb{X} \times \mb{U} \times \mb{W}\to \mb{X}$ and $h:\mb{X} \times \mb{U}  \to \mb{R}^{\dny}$ are bounded functions, with $h(\mb{X},\mb{U})\oplus \mathbb{W} \subseteq \mathbb{Y}$, where $\oplus$ is the Minkowski addition on sets.
The noise structure in \eqref{eq:datagen_sys} corresponds to an \emph{innovation}-type noise process \cite{Cox2021,Wingerden09,Katayama2005, Ljung1999}, which has proven to be rather general for various system classes and has been a cornerstone of the sub-space identification machinery widely used in practice \cite{Katayama2005}. \chris{Finally, we assume that the one-step-ahead predictor of \eqref{eq:datagen_sys} is stable.}

Based on \eqref{eq:datagen_sys}, we have $w_k = y_k-h(x_k,u_k)$, which allows to write the state-equation in the filter form, i.e., 
\begin{equation} \label{eq:filterform|}
 x_{k+1} = f(x_k,u_k,y_k\!-\!h(x_k,u_k)),
\end{equation}
expressing the state evolution without $w_k$.
Based on an observed \chris{length-$N$} data sequence $\mc{D}_{N}=\{(y_k,u_k)\}_{k=1}^N$ generated by \eqref{eq:datagen_sys}, our objective is to identify an LPV model of  \eqref{eq:datagen_sys} in the form of 
\begin{equation}\label{eq:LPV_sys}
\Sigma_{\mr{LPV}} \ : \left\{%
    \begin{aligned} 
        {x}_{k+1} & = {A}(p_k) {x}_k + {B}(p_k)u_k + {K}(p_k)w_k , \\ 
           {y}_k & = {C}(p_k) {x}_k + {D}(p_k)u_k+w_k,
    \end{aligned}\right.
\end{equation}
where $p_k\in\mb{P}\subseteq\mb{R}^{\dnp}$ is the scheduling variable, which varies in the \emph{scheduling space} $\mb{P}$, 
while $A,B,C,D,K$ are bounded matrix functions of $p_k$ with appropriate dimensions. 

The main advantage of modeling \eqref{eq:datagen_sys} in the form of \eqref{eq:LPV_sys} is that it expresses the underlying NL system behavior in a linear form. This allows the use of powerful convex control and observer synthesis methods with stability and performance guarantees \cite{ref:mybook, HoWe15}. This representation capability is achieved via the $p_k$ induced variation of  $A,\ldots,K$. For this purpose, the scheduling variable $p_k$ is considered to be a measurable signal, which is assumed to be a free, independent variable during analysis or controller synthesis for \eqref{eq:LPV_sys}. However, in terms of an LPV embedding of the NL system $\Sigma$, $p_k$ has an underlying connection to the signals in the system with a so-called \emph{scheduling map} $\phi:\mb{X} \times \mb{U} \times \mb{Y}\to\mb{P}$ in terms of \eqref{eq:filterform|}, such that 
the behaviors, i.e., solution sets, of $\Sigma$ and $\Sigma_{\mr{LPV}}$ are equivalent. 

In such an embedding process, a tradeoff in dividing the complexity of $f$ and $h$ in \eqref{eq:datagen_sys} over the dependency structure of ${A},\dots, {K}$ on $p_k$, and the scheduling map $\phi$ is required. To ensure applicability of a wide range of LPV control synthesis approaches on \eqref{eq:LPV_sys}, it is often desired to restrict the matrix functions ${A},\dots, {K}$ to have \emph{affine dependency} on $p_k=[p_{k,1}\, ... \, p_{k,\dnp} ]^\top$, which corresponds to
\begin{equation}\label{eq:affinedep}
A(p_k)= A_0 + {\textstyle\sum_{i=1}^{\dnp}}A_i \,p_{k,i}.
\end{equation}
Hence, it is a highly important problem, even in case when \eqref{eq:datagen_sys} is fully known, to determine a low dimensional scheduling map $\phi$, which allows to represent \eqref{eq:datagen_sys} in the form of \eqref{eq:LPV_sys} with affine dependency of $A,\ldots,K$. As explained in Section \ref{sec:introduction}, \chris{in case of an unknown system \eqref{eq:datagen_sys} with no structural information on $f$ and $h$, there is currently no identification method that can \emph{simultaneously} determine $A,\ldots,K$ and $\phi$, by reaching an optimal complexity-tradeoff, while also dealing with the general innovation-type noise structure.}


To formulate the exact modeling problem of our joint LPV identification problem, 
consider the model structure, i.e.,  the parametrized \emph{predictor form} of \eqref{eq:LPV_sys}, see \cite{Cox2021}: 
\begin{subequations}
 \label{eq:LPV:mod}
\begin{align} \label{eq:LPV:mod:a}
        \hat{x}_{k+1} & ={A}_\theta(\hat p_k)  \hat{x}_k +  {B}_\theta( \hat p_k) u_k + {K}_\theta( \hat p_k) \hat{e}_k, \\ 
           \hat{y}_k & = {C}_\theta(\hat p_k) \hat{x}_k + {D}_\theta(\hat p_k)u_k, \label{eq:LPV:mod:b}
           \end{align}
with $A_\theta,\ldots, K_\theta$ being affine functions of $p_k$, $\theta\in\Theta \subseteq \mathbb{R}^{n_\theta}$ is the collected vectorized form of all matrices in these matrix functions, constituting the parameters of \eqref{eq:LPV:mod:a}-\eqref{eq:LPV:mod:b}, and $\hat{e}_k=y_k-\hat{y}_k$ is the estimate of the innovation noise process $w_k$. Furthermore, \chris{due to the general innovation-type noise structure, we will need to introduce a \emph{partitioned} form of the scheduling map that provides an estimate $\hat{p}_k$ of the scheduling signal $p_k$, i.e.,}
\begin{equation}  \label{eq:LPV:mod:c}
\hat{p}_k= \phi_\eta (\hat{x}_k,u_k,\chris{y_k):=\begin{bmatrix} \phi^\mr{x}_\eta(\hat{x}_k,u_k,y_k)^\top \!&\! \phi^\mr{y}_\eta(\hat{x}_k,u_k)^\top \end{bmatrix}^\top\!,}
\end{equation}
\chris{where \eqref{eq:LPV:mod:a} is only dependent on $\phi^\mr{x}_\eta$ and \eqref{eq:LPV:mod:b} is only dependent on $\phi^\mr{y}_\eta$. The scheduling map $\phi_\eta$ is parametrized with the parameters $\eta \in \Omega\subseteq\mathbb{R}^{n_\eta}$.}
\end{subequations}
Note that \eqref{eq:LPV:mod} corresponds to a joint LPV-SS model-structure, where the state-space matrices and the scheduling map are both included and parametrized, respectively. For a given choice of $\theta\in\Theta$ and $\eta \in \Omega$,  $\hat{e}_k$ corresponds to the \emph{one-step-ahead prediction error} of \eqref{eq:LPV:mod}. Furthermore, if we assume that there exists a  $\theta_0\in\Theta$ and $\eta_0 \in \Omega$ such that $A_{\theta_0},\ldots, K_{\theta_0}$ and $\phi_{\eta_0}$ are equivalent with the corresponding functions in the LPV embedding \eqref{eq:LPV_sys} of \eqref{eq:datagen_sys}, i.e., \eqref{eq:datagen_sys} is \emph{identifiable} under the considered parametrization in \eqref{eq:LPV:mod}, then $\hat{y}_k$ under $(\theta_0,\eta_0)$ is equal to the conditional expectation of $y_k$ w.r.t. $u_k$ and past measurements $\{u_l,y_l\}_{l=-\infty}^{k-1}$. Hence, minimization of the variance of $\hat{e}_k$ in terms of $(\theta,\eta)$ under the data set $\mc{D}_{N}$ 
generated by \eqref{eq:datagen_sys}, i.e., minimization of the \emph{identification criterion} (cost function)%
\begin{equation}\label{id:cost}
	V_{\mathcal{D}_{N}}^{\text{pred}}(\theta,\eta) = \tfrac{1}{N} {\textstyle \sum_{{k=1}}^{N} }\left \| y_{k}  - \hat{y}_{k}\right \|^2_2, 
\end{equation}
corresponds to finding the best LPV model estimate in the conditional expectation sense. This \chris{is in line with} the widely used PEM concept in the LPV and NL identification literatures \cite{toth2012prediction,Sjoberg1995} and has been the cornerstone of many identification methods that are successfully used in practice. However, minimization of \eqref{id:cost} in $\theta$  alone is a highly challenging NL optimization problem, which is coupled with the function estimation problem of $\phi_\eta$. This requires a flexible, yet compact parametrization of $\phi_\eta$ that enables affine representation of matrix variations in \eqref{eq:LPV:mod}. Furthermore, minimization of \eqref{id:cost} also requires estimation of the initial state $\hat{x}(1)$ of the model \eqref{eq:LPV:mod}, which can have a great influence on the \chris{stochastically efficient overall estimate \cite{Toth22a}}. 
In this paper, we develop a novel ANN-based estimator to efficiently tackle the joint LPV identification problem, through the minimization of the prediction error \eqref{id:cost} in terms of \eqref{eq:LPV:mod}.

\shortsection{LPV Sub-Space Encoder Network}\label{sec:preliminaries}
\noindent We propose to use the core idea of SUBNET, 
to derive a \chris{novel} deep-learning-based LPV system identification method. 
By our knowledge, this is the first global LPV identification method that is capable \chris{of simultaneous estimation of $A,\dots,K$ and $\phi$.} 
The proposed LPV variant of SUBNET builds on two main ingredients: a truncated prediction-loss-based cost function and a sub-space encoder, which is linked to the concept of state reconstructability.

\shortsubsection{Batch prediction loss} 
\noindent Consider the scheduling map $\phi_\eta$  as a multi-layer ANN, parametrized in $\eta\in\Omega$, where each hidden layer is composed from $n_\sigma$ activation functions $\sigma:\mathbb{R} \rightarrow \mathbb{R}$ (e.g., sigmoid, ReLU, etc., see \cite{ref:bookonDeepLearning}) in the form of $z_{i,j} = \sigma(\sum_{j=1}^{n_\sigma}\eta_{\mathrm{w},i,j} z_{i-1,j}+ \eta_{\mathrm{b},i,j})$, where $z_i=\mathrm{col}(z_{i,1},\ldots,z_{i,n_\sigma})$ is the latent variable representing the output of layer $1\leq i\leq q$. For $\phi_{\eta}$ with $q$ hidden-layers and linear input and output layers, this means $\phi_{\eta}(\hat{x}_{k}, u_{k}, \chris{y_{k}})= \theta_{\mathrm{w},q+1} z_q(k) + \theta_{\mathrm{b},q+1}$     and $z_{0}(k)=\mathrm{col}(\hat{x}_{k}, u_{k}, \chris{y_{k}})$. With such a form of $\phi_\eta$ and linear parametrization of \eqref{eq:LPV_sys} in $\theta$ it is not very difficult to formulate the minimization problem of \eqref{id:cost} in popular software packages, such as PyTorch, and provide deep-learning-based joint estimates. 

However, the first obstacle to successfully solve the identification problem in this way is that the computational cost of \eqref{id:cost}  scales at least linearly with $N$. Furthermore, despite the advanced \emph{automatic gradient calculation} tools and \emph{stochastic gradient descent} (SGD) methods \cite{ref:bookonDeepLearning}, the forward iterated prediction cost function  \eqref{id:cost} is sensitive to local minima, and gradient-based optimization methods commonly display unstable behaviors~\cite{ribeiro2020smoothness}. To avoid this problem, an important observation is  that \eqref{id:cost} can be well-represented as a sum of prediction subsections of truncation length $T$:
\begin{subequations}
\label{eq:encoder}
\begin{align}
    V_{\mathcal{D}_{N}}^{\text{trun}}(\theta,\eta) &\! =\! \frac{1}{T(N\!-\!T\!+\!1)}\!\!\!\!\! \sum_{{t=1}}^{N-T+1} \sum_{{k=0}}^{T-1}\! \|\hat{y}_{t+k\mid t} \!-\! y_{t + k}\|^2_2, \label{eq:encoder:cost}\\
    \hat{x}_{t+k+1\mid t} & ={A}_\theta(\hat p_{t+k\mid t})  \hat{x}_{t+k\mid t} +  {B}_\theta( \hat p_{t+k\mid t}) u_{t+k}  \notag\\ &\quad + {K}_\theta( \hat p_{t+k\mid t}) \hat{e}_{t+k\mid t}, \label{eq:encoder:b} \\
     \hat{y}_{t+k\mid t} & = {C}_\theta(\hat p_{t+k\mid t}) \hat{x}_{t+k\mid t} + {D}_\theta(\hat p_{t+k\mid t})u_{t+k}, \label{eq:encoder:c} \\
    \hat{e}_{t+k\mid t} &= y_{t+k} - \hat{y}_{t+k\mid t}, \label{eq:encoder:d} \\
    \hat{p}_{t+k \mid t} &= \phi_\eta (\hat{x}_{t+k \mid t},u_{t+k},\chris{y_{t+k}}), \label{eq:encoder:e}
\end{align}
\end{subequations}
where the   notation $|$ is introduced to distinguish between subsections as $(\text{current index} | \text{start index})$. If the truncation length is set to $T = N$, then \eqref{id:cost} is recovered. A significant advantage of \eqref{eq:encoder} is that it increases the loss function smoothness~\cite{ribeiro2020smoothness}, making gradient-based optimization methods more stable and 
{consistent}. 
Moreover, the computational cost of \eqref{eq:encoder} can be reduced from $\mathcal{O}(N)$ to $\mathcal{O}(T)$ by using modern hardware (e.g., GPU architectures), as the sum over $t$ in \eqref{eq:encoder:cost} can be computed in parallel. 

The computational cost of \eqref{eq:encoder} can be further decreased by not summing over all the subsections of the complete data set $\mathcal{D}_N$ for each optimization step, but only over a subset of subsections. This results in a \emph{batch-$\ell_2$-loss} formulation:
\begin{equation}
\label{eq:enc:loss}
    V_{\mathcal{D}_{N}}^{(\text{batch})}(\theta,\eta) = \frac{1}{T\cdot N_{\text{batch}}} \sum_{{t} \in \mathcal{I}}  \sum_{{k=0}}^{T-1} \|\hat{y}_{t+k\mid t} - y_{t + k}\|^2_2 
\end{equation}
where $ \mathcal{I} \subset \mathbb{I}_1^{N-T+1}=\{1,2,...,N-T+1\} $ contains a selection of possible batches in \eqref{eq:encoder} and $N_{\text{batch}} =|\mathcal{I}| $. \eqref{eq:enc:loss} allows the use of powerful batch SGD algorithms such as the Adam optimizer~\cite{kingma2014adam}. 
Despite the computational advantage, an important shortcoming of minimization of \eqref{eq:enc:loss} is that the initial state $\hat{x}_{t\mid t}$ of each subsection is unknown. Instead of introducing a free parametrization of $\hat{x}_{t\mid t}$, which would quickly result in an explosion of the computational costs and the model variance, 
\chris{we now propose} an encoder-based state estimator that uses the concept of state reconstructability, \chris{which is an LPV variant of the core idea of SUBNET.} 


\shortsubsection{LPV sub-space encoder} 
\noindent To derive a state-estimator for the batches in \eqref{eq:enc:loss}, note that the future output evolution of \eqref{eq:LPV_sys} is
\begin{subequations}\label{eq:obsv:01}
\begin{align}
	y_k = & \, {C}(p_k) {x}_k + {D}(p_k)u_k+w_k, \\[-2mm]
	\vdots \> & \notag \\[-4mm]
	y_{k+n} = & \, {C}(p_{k+n})\! \prod_{l=0}^{n-1}\!\tilde{A}(p_{k+l}) {x}_k + {D}(p_{k+n})u_{k+n} +w_{k+n} \notag \\
	& + {C}(p_{k+n})\! \sum_{\tau=0}^{n-1}\! \prod_{l=1}^{n-1}\! \tilde{A}(p_{k+l+\tau})  \tilde B(p_{k+\tau})u_{k+\tau} \notag \\
	& + {C}(p_{k+n})\! \sum_{\tau=0}^{n-1}\! \prod_{l=1}^{n-1}\! \tilde {A}(p_{k+l+\tau})   K(p_{k+\tau})\chris{y_{k+\tau}},
\end{align}
\end{subequations}
where $\tilde{A}=A-KC$ and $\tilde{B}=B-KD$. By introducing $u_{k}^{k+n}\!=\! [\begin{array}{ccc} u_k^\top & \cdots & u_{k+n}^\top \end{array}]^\top$ with $y_{k}^{k+n}$ and $p_{k}^{k+n}$ similarly defined, \eqref{eq:obsv:01} can be written more compactly as
\begin{equation} \label{eq:obsv:02}
 (\mathcal{E}_n \diamond p)_k {y}_{k}^{k+n}= (\mathcal{O}_n \diamond p)_k
{x}_k  + (\mathcal{L}_n \diamond p)_k  u_{k}^{k+n} + w_{k}^{k+n}
\end{equation}
where $\mathcal{O}_n$ is the $n$-step observability matrix of \eqref{eq:LPV_sys},  $\mathcal{L}_n$ is the input transition matrix with lower triangular Toeplitz structure,  $\mathcal{E}_n$ is the output filtering matrix (can be seen as part of an extended observability matrix) corresponding to the innovation process and $\diamond$ expresses the dynamic polynomial dependence of these matrix functions on $p_{k}^{k+n}$. Under a given scheduling sequence $p_{k}^{k+n}$ (i.e., known scheduling map $\phi$) and structural observability of \eqref{eq:LPV_sys} (i.e., invertibility of $\mathcal{O}_{n_\mathrm{x}}$): \vspace{-3mm}
\begin{equation}
 {x}_k \! =\!(\mathcal{O}_n \diamond p)_k^{-1}\! \Big( (\mathcal{E}_n \diamond p)_k {y}_{k}^{k+n} \!\!-\! (\mathcal{L}_n \diamond p)_k  u_{k}^{k+n} \!\!-\! w_{k}^{k+n} \Big )
\end{equation}
has been exploited in LPV sub-space identification to provide estimates of $A,\ldots,K$ \cite{Cox2021,Wingerden09}. However, in case of the considered joint estimation problem, $\phi$ is unknown and it is potentially dependent on the state ${x}_k$, requiring to consider \eqref{eq:obsv:02} as an NL observability problem in ${x}_k$ \cite{Is95}. Based on the back-substitution of $\phi$, such a problem can be solved, giving the $n$-step observability map $ {x}_k = \Lambda_{n}(u_{k}^{k+n},y_{k}^{k+n},w_{k}^{k+n})$ under $n\geq n_\mathrm{x}$. Similarly, an $n$-step reconstructability map $ {x}_k = \Psi_{n}(u_{k-n}^{k},y_{k-n}^{k},w_{k-n}^{k}) $ can be established, by considering the reconstruction of ${x}_k$ from past samples of $u_k, y_k, w_k$.  For the exact construction and the required conditions, see \cite{Toth22a}. 
Note that in practice, the noise sequence $w_{k-n}^{k}$ is not directly available to compute this recovery based on $\Psi_n$. We can however exploit the i.i.d.~white noise property of  $w_k$ to arrive at an efficient estimator $\bar{x}_k$ of ${x}_k$, given by
\begin{equation} \label{eq:state:est}
\bar{x}_k=\mathbb{E}_w\{x_k \mid u_{k-n}^{k}, y_{k-n}^{k} \} = \bar{\Psi}_n(u_{k-n}^{k}, y_{k-n}^{k}).
\end{equation}

Considering \eqref{eq:LPV:mod}, we can similarly use $\bar{\Psi}_n$ to provide an estimate of $\hat{x}_{t\mid t}$ for initialization of each batch computation in \eqref{eq:enc:loss} from past IO data. However, the exact calculation of this estimator for a given ANN parametrization of the scheduling map $\phi_\eta$ and $A_\theta,\ldots,K_\theta$ is practically infeasible. This is due to the required analytic inversion for $\Psi_n$, and the computation of the conditional expectation of $\Psi_n$ under an unknown $\Gamma_\mathrm{w}$. 
Hence, similar to \cite{Toth22a}, we aim to approximate $\bar{\Psi}_n$ by introducing a NL \emph{sub-space encoder} function: 
\begin{equation} \label{eq:sub:enc}
    \hat{x}_{t\mid t} \triangleq \psi_\xi(u_{t-n}^{t}, y_{t-n}^{t}). \\
\end{equation}
Here, $n$ corresponds to the number of past inputs and outputs,

\noindent i.e., \emph{lag window}, considered to estimate the initial state  $\hat{x}_{t\mid t}$, while $\xi\in\Xi\subseteq \mathbb{R}^{n_\xi}$ is the collection of the parameters associated with $\psi_{\xi}$ in terms of a corresponding ANN with multiple hidden layers. In order to provide an estimator for the initial state of the considered model structure \eqref{eq:LPV:mod}, the encoder function $\psi_{\xi}$ is co-estimated with the scheduling map $\phi_\eta$ and the matrices in $A_\theta,\ldots, K_\theta$ by including the parameters $\xi$ in the loss function \eqref{eq:enc:loss}.

\begin{figure}[t]
\centering
\includegraphics[width=\linewidth]{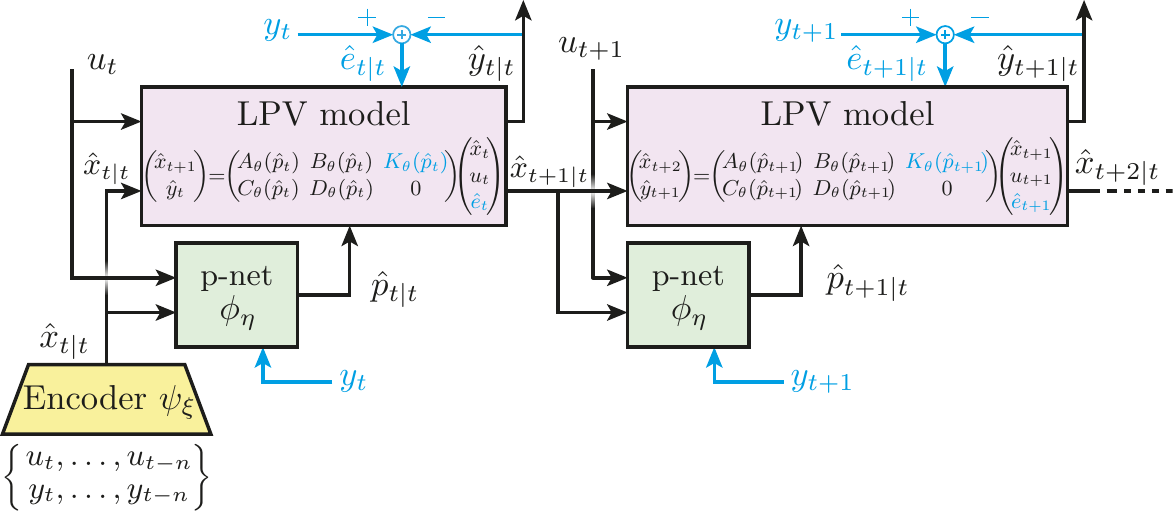} \vspace{-7mm}
\caption{The deep-learning based LPV system identification approach \methodname with self-scheduling.}\label{fig:single-use} \vspace{-5mm}
\end{figure}

\shortsubsection{Network structure of the estimator} 
\noindent Fig.~\ref{fig:single-use} summarizes the ANN structure of our proposed method, which allows to minimize the batch $\ell_2$ loss \eqref{eq:enc:loss} by exploiting the derived sub-space encoder \eqref{eq:sub:enc} and the forward propagation model structure in \eqref{eq:encoder:b}-\eqref{eq:encoder:e}. This structure, named as \methodname, cleverly embeds the structural LPV information in SUBNET.
The \methodname consists of a (deep) forward ANN-based encoder $\psi_\xi$ that takes as input the last $n$ measurements of $u$ and $y$ from the data-generating system. From this data, the encoder estimates the model state at time-step $t$, which in turn is forward propagated by the $\theta$-parametrized LPV model equations in terms of \eqref{eq:encoder:b}-\eqref{eq:encoder:c} and the innovation \eqref{eq:encoder:d}. The required scheduling sequence is determined by a (deep) forward ANN scheduling map $\phi_\eta$, called p-net, that uses the previously calculated $\hat{x}_{t+k\mid t}$ to compute the scheduling for the next state update. This enforces that the LPV-SS model state can be used to determine the scheduling, which is often called \emph{self-scheduling} in the LPV literature. Note that  the structure in Fig.~\ref{fig:single-use} is formulated under the general innovation-type noise model. In case the expected noise structure in the system is OE, i.e., $K(p_k)\equiv 0$ in the LPV embedding of the original system, the network structure can be simplified by dropping the parts highlighted in blue. Furthermore, we only depicted the first two steps in time in Fig.~\ref{fig:single-use} for the batch calculation, while in fact \chris{$T$} steps are considered in the algorithm.

\begin{figure}[t]
\centering
\includegraphics[width=\linewidth]{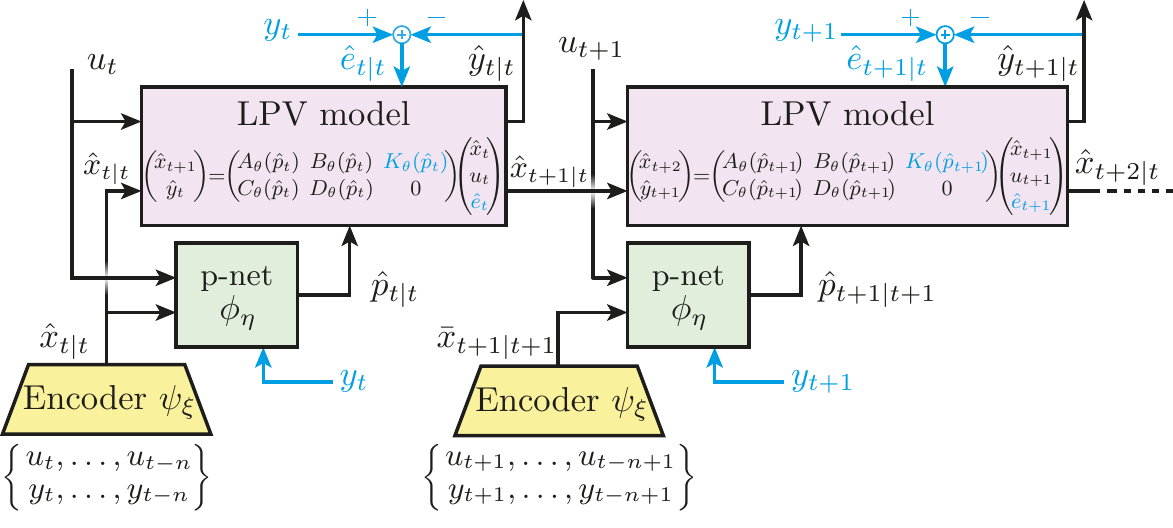} \vspace{-7mm}
\caption{The deep-learning based LPV system identification approach \methodname with external scheduling. \chris{Note that $\psi_\xi$ is the same for every time-step, but fed with a shifted data-sequence.}}\label{fig:multi-use} \vspace{-5.5mm}
\end{figure}

Alternatively, the encoder $\psi_{\xi}$ can be used to estimate the possibly required $\hat{x}_k$ in each time-step, separating the scheduling map calculation from the forward propagation of the model. This formulation considers the scheduling as an external signal determined by a filter operation directly from the data-generating system, which is in line with the intended use of the model for analysis and control purposes. The corresponding \methodname structure with external scheduling estimation is depicted in Fig.~\ref{fig:multi-use} for the first two calculation steps in a batch.

\shortsubsection{Parameter estimation and properties} 
\noindent The \methodname approach can be easily implemented in open-source software packages that support the training of neural networks, e.g., PyTorch. In particular, the Adam~\cite{kingma2014adam} optimizer with default parameters provides an efficient training method for the considered cost function and network structure. The original \mbox{SUBNET} method corresponds to a gradient-based NL SS model estimation method under a PEM-like criterion. It has been shown to provide consistency guarantees with the assumption of persistently exciting data and convergence of the optimization to the global minimum of \eqref{eq:enc:loss}, see \cite{ljung1978convergence,Toth22a} for details. Moreover, as \methodname is based on a rather specialized structure of \mbox{SUBNET}, the consistency properties of the SUBNET method are inherited. Note that while the approach provides a rather automated estimation of LPV models, certain hyper-parameters of the model structure, such as the state-order $n_\mathrm{x}$, the lag-window $n$ and the scheduling dimension $n_\mathrm{p}$ are still needed to be either selected by the user or optimized by popular hyper-parameter optimization methods (e.g., Bayesian optimization based approaches). \chris{The dependency structure of $A,\dots,K$ (e.g., affine, polynomial, etc.) is mainly a user-based design choice that depends on the further utilization of the model, e.g., analysis, optimal LPV controller synthesis, etc.}

\begin{figure}[t]
\centering
\includegraphics[width=0.3\linewidth]{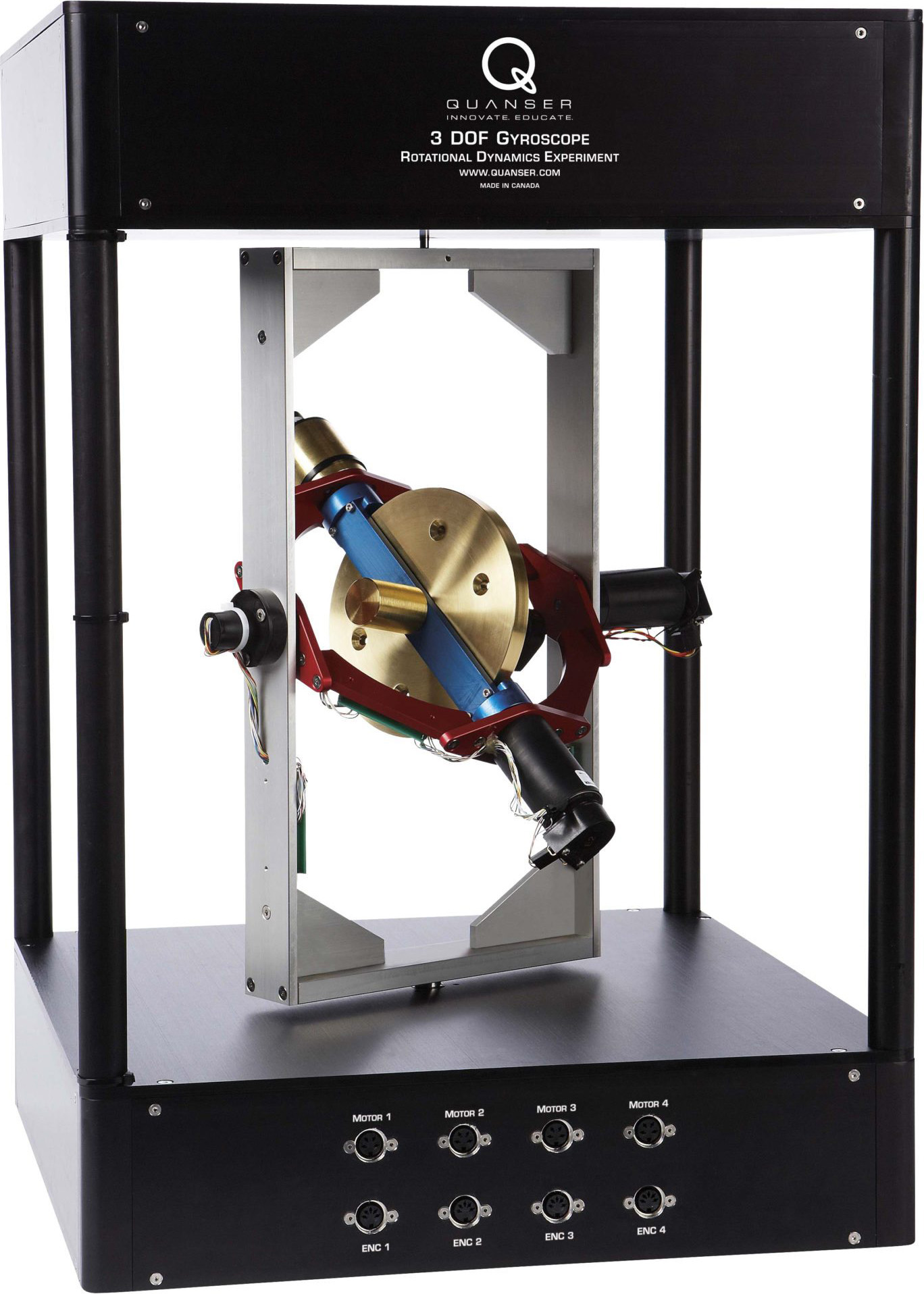}
\vspace{-5pt}
\caption{The CMG system, whose 
simulation model is used for identification
.} 
\vspace{-17pt}
\label{fig:gyro}
\end{figure}

\shortsection{Simulation results}\label{sec:examples}
\noindent In this section, we study the estimation performance of our \methodname 
method on the identification problem of a simulated control moment gyroscope system and compare it with the current state-of-the-art LPV identification methods that use an a priori given scheduling map. \chris{The data sets and code is} available at https://tinyurl.com/lpvsubnet.
\shortsubsection{System description}
\noindent We consider a high-fidelity NL simulator of the 3 degrees-of-freedom \emph{Control Moment Gyroscope} (CMG) setup, depicted in Fig.~\ref{fig:gyro}, for LPV identification. This system consists of a flywheel (with angular position $q_1$), mounted inside an inner blue gimbal (with angular position $q_2$), which in turn is mounted inside an outer red gimbal (with angular position $q_3$). The entire structure is supported by a rectangular silver frame (with angular position $q_4$) that can rotate around its vertical axis of symmetry. The gimbals can rotate freely. The setup is equipped with four DC motors and encoders, actuating and measuring the position of the flywheel and the gimbals. 
In this study, we consider the high-fidelity simulation model of the CMG as the data-generating system, which is a highly-complex and NL system due to the involved rotational dynamics. We consider the following scenario:
\begin{itemize}
\item $q_3=0$  is locked,  $q_2$ is actuated with motor current $i_2$ as an input, while we measure $\dot{q}_4$ as an output.
\item $\dot{q}_1$ is controlled independently with a random multi-level-reference signal with amplitude in $[30,50]$ rad/s and a dwell-time between 4 and 8 seconds.
\end{itemize}
Simulation of the CMG is done in \emph{continuous time} (CT) with fixed-step RK4, sub-sampled to 100 Hz to obtain our data sets. By analyzing the first-principles based dynamics (see, e.g., \cite{bloemers2019equations}), it can be noted that, due to the independent control of the flywheel and the locked status of $q_3$, $\dot{q}_1$ can be regarded as an external input to a model describing the motion of $q_2$ and $q_4$. Finally, we want to emphasize here that the LPV identification methods with which we compare our learning approach, \emph{require} a scheduling map definition. Analyzing a possible global LPV embedding of the DT CMG dynamics yields a scheduling map that is defined as
\begin{equation}\label{eq:schedvar}
p := \phi(x,u) = \begin{bmatrix} \dot{q}_1 & \sin(q_2) & \cos(q_2)\end{bmatrix}^\top.
\end{equation}
We will now compare state-of-the-art LPV identification methods under the choice of \eqref{eq:schedvar} on the CMG with our proposed \methodname method, which is capable of the joint estimation of $\phi$ and an affine LPV-SS model.

\begin{figure}
\centering
\includegraphics[scale=1]{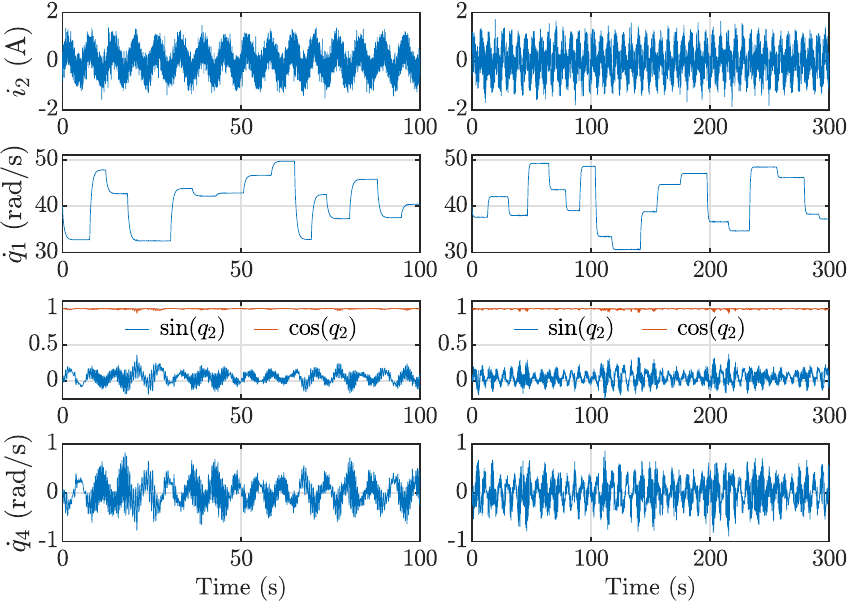}
\vspace{-20pt}
\caption{Estimation (left) and validation (right) data sets from the CMG.}
\vspace{-16pt}
\label{fig:est_val}
\end{figure}

\shortsubsection{Experiment design}
\label{sec:exp:design}
\noindent We excite the 
CMG with the 
current signal:
\begin{equation}
i_{2,k}:= u_k=\tfrac{1}{2} \sin(\omega T_\mathrm{s} k) + v_k
\end{equation}
where $v$ is white with $v_k\sim \mathcal{N}(0,\sigma_\mathrm{v}^2)$ and $\sigma_\mathrm{v}=\frac{1}{3}$ together with $\omega \sim \mathcal{U}(1,2)$ per data set realization. This gives a sinusoidal carrier signal with a frequency in $[1,2]$ Hz and a superimposed white noise excitation with a total magnitude of $1.5$ and a confidence level of $99.7\%$.
Moreover, we consider an OE noise structure. The white noise signal $w$ is such that $w_k\sim \mathcal{N}(0,\sigma_\mathrm{e}^2) $ with variance $\sigma_\mathrm{e}^2=2.2 \cdot 10^{-5}$, which corresponds to a \emph{signal to noise ratio} (SNR) of 35 dB. In order to capture the variation along a wide operating range of the CMG, the estimation data set is of size $N_\mr{d}=10^4$, while the validation data set is of size $N_\mr{d}=3\cdot 10^4$. \chris{The validation set is uncorrelated to the estimation set}. The resulting data sets are depicted in Fig.~\ref{fig:est_val}. 
For the \methodname method, we consider the measured $i_2$ and $\dot{q}_1$ as the inputs and $\dot{q}_4$ as the output in terms of an available data set, while for regular LPV identification methods we use only $i_2$ as input, $\dot{q}_4$ as output and \eqref{eq:schedvar} as scheduling signal, providing significant structural information. 
\shortsubsection{Estimation of LPV models}
\noindent We compare our approach to the current state-of-the-art global LPV system identification methods. Particularly, LPV-OE \cite{toth2012prediction, zhao2012prediction} 
and LPV PEM-SS \cite{Toth18cAUT, Cox2021} identification. Only the considered hyper-parameters and initialization methods are discussed, while the details on these approaches are omitted. For the execution of the identification procedures, we use the \matlab toolbox \lpvcore
\cite{lpvcore}.

\subsubsection{LPV-OE identification}
This approach aims to estimate the system in the form of an LPV-IO model with OE noise structure \chris{(i.e., $y_k = \hat y_k+ e_k$)}:
\begin{equation} \label{lpv:Oe}
\hat{y}_k + {\textstyle \sum_{i=1}^{\dna}}a_i(p_{k-i})\hat{y}_{k-i}  = {\textstyle\sum_{j=0}^{\dnb}}b_j(p_{k-j})u_{k-j},
\end{equation}
under the cost function \eqref{id:cost}. In \eqref{lpv:Oe},  $\{a_i\}_{i=1}^{\dna}$ and $\{b_j\}_{i=0}^{\dnb}$ correspond to coefficient functions that have shifted affine dependency on $p_k$ \cite{toth2011state}. Note that such a dependency structure ensures a direct minimal SS realization of the estimated models in the form of \eqref{eq:LPV_sys} with affine dependency and $K=0$, see \cite{toth2011state}, making the results comparable with \methodname. 
Based on the given data, an estimate of \eqref{lpv:Oe} is computed with {\tt lpvoe}
in \lpvcore, under $\dna=\dnb=5$. 

\subsubsection{LPV PEM-SS identification} This approach corresponds to a direct gradient based estimation of \eqref{eq:LPV:mod:a}-\eqref{eq:LPV:mod:b} via the minimization of \eqref{id:cost}, often initialized via sub-space or OE estimation. A model estimate under $n_\mathrm{x}=5$ has been computed by {\tt lpvssest} in \lpvcore using default options and an LPV-OE initialization for the sake of consistency. 
%
%

\subsubsection{\methodname identification}
In order to have a fair comparison, we tune the hyper-parameters for the \methodname method similar to the LPV identification methods. Therefore, the output dimension of $\psi_\xi$, i.e., the state-dimension is set to 5, as well as the lag window $n$. The output dimension of the p-net $\phi_\eta$, i.e., the scheduling dimension, is set to 3. The encoder $\psi_\xi$ has 2 hidden layers, each with 64 neurons, $\tanh$ activation functions, and a linear bi-pass to capture linear dependencies. The p-net has the same structure as $\psi_\xi$. The truncation length $T$ is initially set to 5 and has been increased to 60 in the first few epochs to enhance training stability. The batch size during optimization has been 256. We have used the Adam optimizer \cite{kingma2014adam} with the standard configuration settings as implemented in PyTorch.

\begin{figure}
\centering
\includegraphics[scale=1]{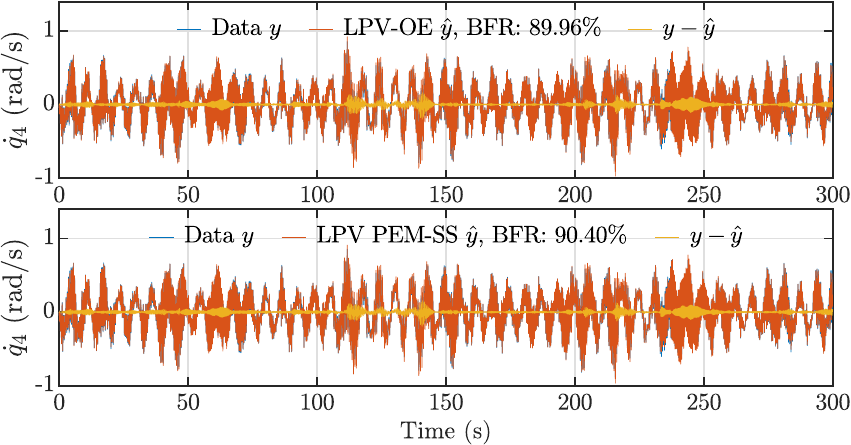}
\vspace*{-20pt}
\caption{Simulation error of the models identified by the LPV identification methods LPV-OE (top) and LPV PEM-SS (bottom) on validation data.}
\vspace{-16pt}
\label{fig:comp_infstep}
\end{figure}

\begin{figure}
\centering
\includegraphics[scale=1]{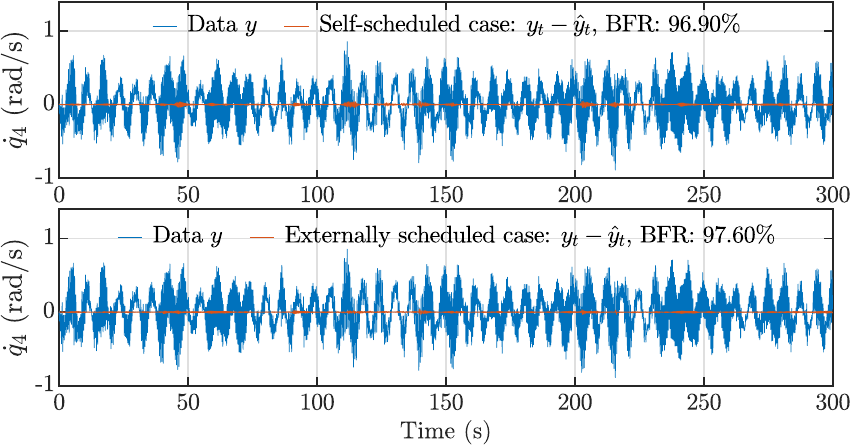}
\vspace{-20pt}
\caption{Simulation error of the models identified by the \methodname method on validation data: self-scheduled LPV model (top) and externally scheduled LPV model (bottom).}
\vspace{-10pt}
\label{fig:learned_res}
\end{figure}

\shortsubsection{Comparison of the results}
\noindent Simulation responses of the estimated models are computed on an \chris{uncorrelated test} data set, generated according to Section \ref{sec:exp:design} with $N=3\cdot 10^4$, and 
are depicted in Fig.~\ref{fig:comp_infstep} and Fig.~\ref{fig:learned_res}, respectively.
Note that due to the OE noise structure, simulation and prediction responses of the estimated models are equivalent. We use the \emph{Best Fit Rate}\footnote{$\mathrm{BFR}=\max\!\big\{\hspace{-0.5mm}1\!-\!\frac{\frac{1}{N}\!\sum_{t=1}^N\!\Vert y_t-\hat{y}_t\Vert_2}{\frac{1}{N}\!\sum_{t=1}^N\!\Vert y_t-\bar{y}\Vert_2}, 0\!\big\} \cdot 100\%$ with $\bar{y}$ being the sample mean of $y$.}
(BFR) in terms of the simulation accuracy as our criterion for the correctness of the result. Under the OE noise structure and the SNR of 35 dB, we can at most achieve a BFR of 98.22\%. 
Based on Fig.~\ref{fig:comp_infstep}, the resulting simulation accuracy of the LPV models shows around 90\% of BFR, which shows that the obtained models using these current state-of-the-art LPV identification methods give a relatively accurate representation of the system behavior with the selected scheduling map \eqref{eq:schedvar}. Computation of these models lasted for a few minutes.

The results for \methodname were obtained by training the network on a consumer grade computer for 10 hours with 200,000 batch updates on the estimation data set and early stopping on the validation data set to avoid over-fitting. 
The resulting simulation accuracy of the trained models shows around 97\% of BFR in Fig.~\ref{fig:learned_res}, which shows that the trained model is an almost perfect \chris{approximation} 
of the underlying NL system. 
Note that this is about 3 times better than the LPV identification results, without requiring a user specified scheduling map. Also note that the same data set has been used for all methods, meaning that the number of data points has been sufficient for both the regular LPV identification methods and also for our \methodname method, despite being a deep-learning-based approach. The latter underlines the efficiency of our method. In terms of utilization of the obtained models, the self-scheduled \methodname model \chris{(Fig.~\ref{fig:single-use})} is interesting as it can self-generate its scheduling signal without needing a direct estimation or measurement of internal signals of the represented system, making it useful for simulation purposes. The externally scheduled \methodname model \chris{(Fig.~\ref{fig:multi-use})} is of particular interest for LPV model-based controller design as it provides a scheduling calculation for controller implementation from direct IO measurements of the NL system. 
\shortsection{Conclusions}\label{sec:conclusion}
\noindent\methodname is the first approach capable of simultaneously estimate an LPV-SS model jointly with scheduling map for a complex NL system, directly from measured input-output data without additional structural information. The resulting scheduling map estimate not only solves the difficult selection procedure for the user, but provides a particularly useful calculation of the scheduling signal, which makes the resulting model directly usable for follow-up analysis and control. \methodname has consistency guarantees under a general innovation noise setting and is easily implementable in open-source software packages. Our simulation results on the CMG showcase that the \methodname can efficiently achieve highly accurate models.

\bibliographystyle{IEEEtran}
\bibliography{ref_cdc2022}

\end{document}